УДК 339.13.017.

## Новая экономика: от кризиса доткомов к виртуальному бизнесу
*Калужский М.Л.*

***Аннотация***: *Статья об объективных закономерностях и механизмах формирования сетевой экономики в условиях кризиса «новой экономики». Сетевая экономика рассматривается как основа следующего экономического цикла. Автор анализирует особенности институциональной трансформации экономических отношений под влиянием развития облачных технологий и виртуальных организаций.*
***Ключевые слова***: *дот-ком, новая экономика, сетевая экономика, институционализм, интернет, облачные технологии, виртуальная организация.*

## New economy: from crisis of dot-coms to virtual business
*Kaluzhsky M.L.*

***Abstract***: *Article about objective regularity and mechanisms of formation of network economy in the conditions of crisis of «new economy». The network economy is considered as a basis of a following business cycle. The author analyzes features of institutional transformation of economic relations under the influence of development of cloudy technologies and the virtual organizations.*
***Keywords***: *dot-com, new economy, network economy, institutionalism, Internet, cloudy technologies, virtual organization.*

**Предыстория**. Структурные изменения в экономике конца XX – начала XXI веков традиционно ассоциируются с понятием «новая экономика», которое появившись в 1980-х гг., первоначально обозначало экономику постиндустриального общества, ориентированную на сферу услуг. В 1990-х гг. этим термином начали называть сферу экономических отношений, связанных с использованием высоких технологий.

Сегодня под «новой экономикой» понимается «*влияние высоких технологий на экономическое окружение, которое ведёт к изменению отдельных макроэкономических параметров*» [1, с. 161]. К ней традиционно относят биотехнологии, микроэлектронику, интернет-технологии и ряд других направлений, которые формально принадлежат к различным отраслям экономической деятельности. Однако, если в 1990-е гг. новая экономика ассоциировалась с компьютерными технологиями, то на рубеже веков цикл её развития завершился кризисом доткомов.[*]

В пертурбациях последних лет уже мало кто помнит, что первый локальный экономический кризис текущего тысячелетия был связан именно с новой экономикой. Всё началось в середине 1990-х годов, когда резкий взлёт доходности ведущих интернет-компаний создал у инвесторов иллюзию сверхвысокой доходности этого вида бизнеса. В новую экономику хлынули огромные инвестиции, а эмиссия акций крупнейших компаний, связанных с интернет-технологиями, стала приносить гораздо больше прибыли, чем непосредственная экономическая деятельность.

Историческое схлопывание экономического пузыря началось 10 марта 2000 года, когда индекс NASDAQ упал в течение дня с 5132,52 пункта в начале биржевых торгов и до 5048,62 при их закрытии. Затем последовала волна банкротств компьютерных компаний, падение индекса NASDAQ, а также обвал цен на серверные технологии и компьютерные решения.

Россию кризис доткомов по вполне понятным причинам обошёл стороной. У нас не было высококапитализированных компьютерных компаний, чьи акции могли котироваться на мировых биржах. Однако именно кризис доткомов стал первым звоночком, ознаменовавшим начало коренной трансформации мировых экономических отношений.

Строго говоря, кризисы в экономике почти никогда не бывают связаны с исчерпанием потенциала дальнейшего развития. Практически все они являются следствием неадекватного восприятия реальности экономическими субъектами и основанных на этом ошибочных стратегий экономического поведения. Кризис доткомов также не стал исключением, а лишь скорректировал общий вектор мирового экономического развития.

**Особенности кризиса доткомов**. Причины кризиса во многом были вызваны абсолютным непониманием участниками рынка особенностей ведения такого рода бизнеса. Экономические субъекты, не понимая, что имеют дело с виртуальными ценностями, попытались применить к господствующим в новой экономике отношениям стереотипы экономического поведения в старой (материальной) экономике.

Говоря на языке институциональной теории, причиной кризиса стала попытка переноса старых институтов в новые экономические условия. Однако институциональные процессы в новой экономике не просто другие. У них совершенно иной (противоположный) вектор институционального развития: от концентрации экономических преимуществ к их размыванию.

Это обстоятельство стало полной неожиданностью для экономических субъектов, действия которых были адекватны ситуации с точностью до наоборот. Так, если на начальном этапе субъекты новой экономики демонстрировали умопомрачительные показатели рентабельности, то впоследствии ситуация кардинально менялась. Основным отличием новой экономики от привычных экономических отношений стали как минимум два её признака, вызвавших неминуемый кризис:

1. *Убывающая полезность*. Выгоды от использования новых технологий не связаны напрямую с производством материальных благ. Они определяются не внутренними характеристиками продукта, а условиями его применения. Жизненный цикл таких технологий очень короток. Всё, что не окупилось в самой ближайшей перспективе, стремительно устаревает и утрачивает свою ценность. Тогда как получаемые преимущества от использования информационных технологий имеют смысл только в условиях отсутствия таковых у конкурентов.

С другой стороны, стратегии производителей, построенные на постоянном обновлении версий и расширении функционала, тоже со временем перестают давать эффект, поскольку запросы потребителей не успевают за предложением продавцов. Тут есть определённая закономерность: чем сложнее информационная технология, тем уже область её применения и тем меньше связанный с нею платёжеспособный спрос.

2. *Убывающая доходность*. В условиях глобального открытого рынка информационных технологий любые попытки не то что повышать, а даже удержать доходность на фиксированном уровне, неизбежно ведут к кризису. Примером может послужить создание в феврале 1998 года «*Open Source Initiative*» (OSI) – организации, продвигающей открытое программное обеспечение в ответ на агрессивную маркетинговую стратегию компании «*Microsoft*».

Особенность ситуации заключается в том, что многомиллионная аудитория и свободный доступ в Интернет до нуля опускает входной барьер на этом рынке. Что позволяет за счёт масштабов окупить внедренческие затраты любому новому участнику, который окажется в состоянии бросить вызов существующим лидерам. Причём успех гарантирован, если предлагаемый продукт будет бесплатным. Примером может служить триумфальное шествие по миру социальных сетей, буквально «похоронивших» платные почтовые сервисы.

Не удивительно, что в результате действия описанных закономерностей произошло углубление дисбаланса между ожиданиями инвесторов и реальной стоимостью компаний, приведшее к кризису. Например, М. Кастельс отмечает, что уже в 2000 г. инвестиции, связанные с информационными технологиями, составляли в США до 50% всего объёма частных капиталовложений [2, с. 122].

Вместе с тем, в этот период на экономическую ситуацию оказывали огромное влияние причины, связанные не с «новой», а со «старой» экономикой. Так, именно на 1990-е гг. пришёлся пик массового переноса промышленного производства из развитых стран в страны Юго-Восточной Азии, и в первую очередь – в Китай. Экономия на стоимости рабочей силы позволяла производителям получить прибыль, значительно превышающую среднеотраслевые показатели.

Это дало огромный толчок подъёму китайской экономике и началу китайского экономического чуда. Так, только за 1981-2000 гг. на производство пришли дополнительно 288 млн. чел. Ежегодно в КНР создавалось около 8 млн. новых рабочих мест. Японские эксперты в 2001 г. подсчитали, что в КНР стоимость труда в 30 раз ниже, чем в Японии. В результате цена услуг в Китае в 8,44 раза ниже, чем в Японии, а цены на промышленные товары – в 2,49 раза ниже [3, с. 15]. Большую роль в изменении структуры мировой экономики сыграла предложенная западным инвесторам формула: «*доступ к китайскому рынку* (труда) *в обмен на современные технологии*» [3, с. 26].

Следует отметить, что для западных инвесторов значительное снижение производственных издержек отнюдь не сопровождалось столь же значительным снижением розничных цен. Наоборот, этот процесс сопровождался усилением роли финансового сектора и сферы услуг в экономике, а также их глобализацией за счёт непропорционально возросшей нормы торговой прибыли крупных корпораций.

Итогом стала ситуация, когда за счет изменения структуры ценообразования на товары, производимые в Китае и других странах Юго-Восточной Азии, были созданы условия для преимущественного развития новой постиндустриальной экономики. Больше всего от этого выиграл банковский сектор, кредитовавший всех участников торговой цепи, а также финансовый сектор за счёт массового притока свободных ресурсов на фондовые рынки. В результате большинство крупнейших транснациональных корпораций (80% или около 400 из 500) осуществили инвестиции в создание производственных мощностей в Китае [3, с. 24]. При этом следует отметить, что производственные издержки, снижены до своего минимума, перестали играть определяющую роль в экономике и стали учитываться здесь что называется «при прочих равных».

Этому есть экономическое объяснение: до кризиса новой экономики 2000-2001 гг. большая часть прибыли ТНК создавалась уже не на производстве, а в финансовом секторе и торговле. Казалось, что так будет продолжаться вечно. В западной экономической теории даже появилось понятие «интеллектуальная компания», т.е. компания, стремящаяся к отказу от материальных активов. При этом степень «интеллектуальности» определялась соотношением между рыночной капитализацией и стоимостью материальных активов.

Так, например, американский экономист Т.Стюарт, призывая отказываться от реальных активов в пользу интеллектуальных ресурсов, приводил сравнительную характеристику компаний «Microsoft» и «IBM». Согласно этим данным рыночная капитализация компании «Microsoft» составляла в 1996 году 85,5 млрд. долларов против 70,7 млрд. долларов у «IBM». Тогда как стоимость материальных активов за вычетом амортизации у компании «Microsoft» составляла лишь 930 млн. долларов против 16,6 млрд. долларов у «IBM» [4, с. 377].

В этом и заключалась стратегическая ошибка американского подхода, в результате которой материальные активы и производственные технологии перетекли в Китай. Проблема была в том, что «рыночная капитализация» компании зачастую определялась биржевой стоимостью её ценных бумаг и рейтинговыми показателями, являясь, по сути, банальным финансовым пузырём. Тогда как реальная стоимость компаний, основанная на показателях продаж в условиях надувания пузыря, никого не интересовала.

В 2000-2001 гг. пузырь безо всякого внешнего воздействия достиг пределов своего роста и лопнул, что ударило по т.н. «интеллектуальным компаниям», капитализация которых резко снизилась. Тогда как Китай получил в своё распоряжение мощнейший тех-нологический ресурс в виде западных технологий и производственных мощностей для дальнейшего развития.

Успех китайской экономики был закреплён вступлением КНР во Всемирную торговую организацию в 2001 году. Основным результатом этого вступления «*стал опережающий рост экспорта с 249 млрд. долл. в 2001 г. до 1200 млрд. долл. в 2007 г. В итоге положительное сальдо торговли товарами и услугами (превышение экспорта над импортом) стремительно росло: 2001 г. – 23 млрд. долл., 2002 г. – 31, 2003 г. – 23, 2004 г. – 67, 2005 г. – 120, 2006 г. – 180, 2007 г. – 264 млрд. долл.*» [3, с. 18].

Неизбежным следствием перераспределения реального промышленного потенциала стал постепенный переход иностранных производственных мощностей под контроль китайских компаний. Самый типичный пример – покупка у американской корпорации «IBM» в 2004 г. за 1,25 млрд. долларов бизнеса по производству персональных компьютеров крупнейшей китайской компьютерной корпорацией «Lenovo» [3, с. 27-28].

**Последствия кризиса доткомов**. Сегодня уже трудно согласиться с тезисом о том, что «*старые базисные инновации – информационные технологии, компьютеры и Интернет – не генерируют достаточного количества вторичных инноваций, которые обеспечили бы рост продуктивности факторов производства, а новые базисные инновации пока не появились*» [5, с. 11]. Информационные технологии, компьютеры и Интернет сами являются вторичным продуктом от использования технологических возможностей инновационных компаний. Поэтому ждать инноваций следует не от них, а от их использования в качестве инструмента достижения иных экономических целей.

Специфика компьютерных (информационных) технологий такова, что стоимость товара здесь формируется не в сфере производства, а в сфере освоения технологий или в сфере использования товара. Сфера освоения технологий определяет внедренческие затраты на разработку и освоение производства новых товаров. Сфера использования товара определяет ту часть будущей прибыли, которой покупатели готовы будут поделиться с производителем товара при его покупке.

При этом сфера использования товара первичная по отношению к сфере освоения технологий. В случае если продукт виртуален или близок к тому, то его рыночная стоимость вообще определяется не производственными издержками (которые минимальны) и не биржевой стоимостью ценных бумаг, а той сравнительной выгодой, которую покупатель получает от его использования.

Жизнь наглядно показала, что все представления о возрастающей доходности компьютерных технологий оказались мифом [4, с. 394]. Клиент заплатит за продукт высокую цену, если впоследствии окупит затраты за счет повышения своей сравнительной конкурентоспособности. Но если клиент приобретает продукт в условиях, когда он уже есть у большинства конкурентов, то потребительская стоимость продукта будет определяться лишь его доступностью. Если же можно использовать нелицензионную копию продукта или альтернативный бесплатный аналог, то потребительская стоимость товара вообще стремится к нулю.

Некоторые маркетологи вообще утверждают, что в новой экономике получать прибыль выше среднеотраслевого уровня могут только компании, занимающие по какому-либо значимому параметру монопольное положение на рынке [6]. Как только монопольное положение на рынке утрачивается, вместе с ним улетучивается и сверхприбыль.

При этом на макроэкономическом уровне глобализация спроса и предложения здесь лишь усиливает в среднесрочной перспективе монопольные тенденции на рынок [7, с. 352]. Следует отметить, что вопреки постулатам общей экономической теории, монополисты в новой экономике ведут себя «как совершенные конкуренты» [1, с. 167-168].

Объясняется такое поведение очень просто: давление на монополистов оказывают не столько (большей частью потенциальные) конкуренты, сколько сам потребительский рынок. Единственно возможная стратегия в таких условиях – стратегия «снятия сливок», ко-гда цена на новый товар сначала запредельно завышается, а затем поэтапно снижается по мере отработки ценовых сегментов рынка.

В качестве примера можно привести различающиеся почти в 30 раз предельные оптовые цены на микропроцессоры компании «Intel» в компьютерном супермаркете «Никс» (по состоянию на 03.08.2012):[**]

CPU Intel Core i7-3960X Extreme 3.3ГГц /1.5+15Мб /5ГТ /LGA2011 = 965,0 US$;
CPU Intel Celeron D351 3.2 ГГц /256K /533МГц LGA775 = 32,5 US$.

Стоимость сырья для изготовления обоих микропроцессоров практически одинаковая. Мало того, в течение короткого времени (1-2 года) цена самого дорогого микропроцессора постепенно опускается до цены самого дешёвого. Неизменной остаётся лишь ценовая линейка, верхнее значение которой отражает покупательную способность клиентов в высшей ценовой категории, а нижнее значение – стоковую цену на рынке.

Особенность ситуации заключается ещё и в том, что жизненный цикл компьютерных технологий как товара в несколько раз короче цикла их применения. Так, например, если ассортиментная линейка персональных компьютеров полностью обновляется примерно раз в 1,5-2 года, то срок их эксплуатации в среднем составляет 5-6 лет. Поэтому взлёт продаж новой технологии в период завоевания рынка отнюдь не гарантирует сохранения высоких показателей продаж там даже в условиях монополизма.

Поэтому для новой экономики требуется принципиально иная методология анализа экономической эффективности инноваций. Одним из вариантов может стать, например, полное или частичное игнорирование технических параметров компьютерной техники, когда учёт ведется на основе средневзвешенных показателей, характерных для текущей ассортиментной линейки. В противном случае мы получаем умопомрачительные показатели, как например: «*в течение жизни одного поколения в 1970-1990-е гг. цена компьютеров понизилась более чем в 10 тыс. раз, или в среднем ежегодно на 30-40%*» [8, с. 5].

Следует понимать, что покупатель платит не за компьютер с мегагерцами и гигабайтами, а за возможность решения с его помощью своих насущных задач. Рыночная стоимость компьютера определяется не его внутренней ценностью (металл и пластмасса стоят копейки), а ценностью решений, достигаемых с помощью компьютера. Ценность компьютера становится производной от его желаемости на первых этапах жизненного цикла товара и от его полезности на последующих. У этого продукта нет минимальной базовой стоимости, как, например, у золотых украшений.

Отсюда следует вывод о том, что *главным фактором ценообразования в новой экономике является способность продавца предложить конкурентоспособный продукт для решения проблем клиентов*. Главное здесь – запросы (часто неосознанные) потребителей и только затем – предложение продавца. Причем ценность продукта определяется конъюнктурой предложения на рынке и конъюнктурой спроса. Эти обстоятельства имеют гораздо большее значение, нежели номинальные параметры продукта.

**Рождение сетевой экономики.** Кризис доткомов 2000-2001 гг. ускорил процесс перехода к следующему этапу в развитии новой экономики – сетевой экономике и связанной с нею электронной коммерции. Говоря языком теории систем, в результате бифуркации мировая экономика перешла на более высокий уровень системной самоорганизации. Как отмечает К.Келли: «*компьютерные чипы и коммуникационные сети создали такой сектор экономики, который привел к трансформации всех других её секторов*» [9, с. 5].

Сегодня этому сектору можно дать следующее определение: *Сетевая (электронная, цифровая) экономика – это коммуникационная среда экономической деятельности в сети Интернет, а также формы, методы, инструменты и результаты её реализации*.

Технологической основой появления сетевой экономики послужило не только и не столько развитие компьютерных технологий, сколько появление широкополосного Интернета. Внедрение скоростного интернета привело к значительному сокращению транcакционных издержек и невиданному прежде ускорению операций. Некоторые исследова-тели сегодня прямо указывают на то, что «*прослеживается прямая зависимость влияния скорости развития пропускной способности широкополосных сетей на структурные сдвиги в экономике и появление в ней новых секторов*» [10].

У сетевой экономики, как у качественно новой (системной) формы новой экономики, существует собственное уникальное отличие. Если доткомы отождествлялись с Интернет-маркетингом производителей товаров (работ, услуг), то сетевая экономика трансформировалась в Интернет-маркетинг посредников и пользователей.

Очень показательна в этой связи маркетинговая политика субъектов новой экономики. Так, например, основные рекламные посылы компьютерных компаний 1990-х гг. в сфере «В2В» были ориентированы на стимулирование спроса потребителей технологий. Им постоянно объясняли, что только покупка новейшей модификации продукта способна обеспечить высокий уровень конкурентоспособности на рынке.

В 2000-х гг. ситуация вышла из-под контроля производителей и сегодня они являются всего лишь поставщиками оборудования. Это выразилось не только в значительном снижении трансакционных издержек на обмен и использование информации. Доминирование сферы производства трансформировалось в доминирование сферы распределения.

Произошёл вполне различимый качественный институциональный прорыв – переход экономической инициативы от транснациональных корпораций к потребителям технологической продукции. Здесь мы имеем дело с теми «*вторичными инновациями*», о которых писал академик В.М.Полтерович в статье «*Гипотеза об инновационной паузе и стратегия модернизации*» [5, с. 11]. Институциональная особенность таких инноваций состоит в том, что они действуют вне сферы влияния институтов государственного регулирования и традиционных экономических субъектов.

Сегодня сетевая экономика представляет собой не просто новые методы и формы экономических отношений. Сетевая экономика – это ещё и принципиально новая экономическая среда. При этом «*наибольший эффект роста производительности происходит не при производстве ИКТ* [информационно-коммуникационных технологий] *…, а при использовании ИКТ*» [10].

Сетевая экономика изначально сформировалась в той сфере, которая отличалась наименьшей организованностью, и буквально за несколько лет изменила сущность и структуру экономических отношений. Размеры организации здесь определяются не объёмами производства или продаж, а степенью самостоятельности в осуществлении трансакций.

Теперь, если «*дешевле осуществлять трансакции самой, организация растет, если наблюдается противоположное, она сжимается. Индустриальная эпоха повлияла на внутренние трансакции, но сети снизили ценность централизованного контроля и уменьшили объем дорогостоящих действий административного характера*» [11, с. 622]. Соответственно, коммуникационные возможности частных предпринимателей, небольших и средних фирм существенно сближаются с возможностями крупных корпораций, равно как и охват целевой аудитории. Расстояния уже не имеют значения, а географические границы рынков стираются.

В результате главным параметром рыночного успеха становится доступ участника рынка к сетевой инфраструктуре и уровень её использования. Согласно «сетевому закону» Роберта Меткалфа полезность (ценность) сети для пользователей и для экономики в целом возрастает пропорционально квадрату числа её участников, т.е. экспоненциально.[***] Поэтому развитие сетевых отношений характеризуется двумя ключевыми параметрами:

1) удельным весом ИТ-сектора, т.е. производящими и инфраструктурными компаниями;

2) удельным весом компаний, строящих свои конкурентные стратегии на сокращении трансакционных издержек через использование Интернет-технологий.

Следует отметить и то, что формальные государственные институты по всему миру оказались не готовы к появлению сетевой экономики. В России, например, официальная статистика пока даже не выработала методик оценки происходящих в сетевой экономике процессов [11, с. 622]. Пока становление этой экономики напоминает открытие нового измерения, результаты протекающих в котором процессов влияют на экономическую реальность, но в прежних измерениях не видны.

Огромным достижением сетевой экономики является её доступность. Развитие сетевой экономики нивелирует границы государств и делает виртуальный бизнес доступным для всех желающих. Предприниматель может находиться в ЮАР, России или Прибалтике, а его сотрудники располагаться на Гавайских островах или в Норвегии. Мало того, для сетевых компаний больше не нужны ни огромные офисы, ни формализованные трудовые отношения.

Вместе с тем, объективное сокращение издержек на проведение рыночных трансакций в сети разрушает само экономическое основание для существования больших корпораций. Им не смену приходят более мелкие узкоспециализированные компании, оказывающие услуги одновременно множеству сторонних клиентов.

В сетевой экономике нет смысла и в выполнении всех экономических функций в рамках одной компании. Теперь компания может сосредоточить усилия на выполнении лишь наиболее рентабельных функций, доверив реализацию остальных более профессиональным аутсорсерам.

Такая схема организации бизнеса привела к парадоксальному результату: сбытовая инфраструктура в сети начинает жить своей собственной жизнью. Она больше не привязана к производителям и торговым посредникам. Наоборот, в сети посредники превращаются в аутсорсеров, а производители не контролируют ничего, кроме своих отпускных цен. В результате трансформируется не только рынок. Трансформации подвергаются вековые устои формирования экономических институтов, прежде всего – принципы организация и ведения бизнеса.

Видный американский экономист Р.Коуз, формулируя определение фирмы, указывал, что её институциональный смысл заключается в осуществлении рыночных трансакций при условии, что «*внутрифирменные издержки меньше, чем издержки рыночных трансакций*» [12, с. 12]. В сетевой экономике, наоборот, издержки рыночных трансакций ниже внутрифирменных издержек и это меняет само понятие фирмы. Фирма из обособленной организационной структуры постепенно превращается в виртуальное сообщество единомышленников, основанное на использовании сетевых коммуникаций.

Поэтому сетевая экономика ведет к размыванию также и межфирменных границ. Месторасположение сотрудников здесь вообще не имеет никакого значения. Они могут находиться в разных частях земного шара, синхронно решая задачи своей компании. Равно как и клиенты могут приобретать товары фирмы у виртуальных продавцов вне традиционных сбытовых сетей независимо от своего местонахождения. Уже сейчас сделки с покупателями из Казахстана и Новой Зеландии при почтовой доставке товара отличается для продавца в России или Китае лишь незначительными различиями в сроках доставки товара.

**От облаков к виртуальному бизнесу**. Дальнейшие перспективы развития сетевой экономики неразрывно связаны с расширением использования облачных вычислений (*cloud computing*) и обусловленных ими виртуальных бизнес-технологий. Экономическая сущность облачных вычислений заключается в использовании общедоступного сетевого доступа к пулу вычислительных ресурсов (серверам, сервисам и приложениям), предоставляемого в сети внешними провайдерами. Институциональная сущность облачных вычислений заключается в экономии на трансакционных издержках, связанных с инфраструктурой информационных технологий.

Вообще, облачные вычисления появились довольно давно. Технология виртуализации была предложена корпорацией «IBM» ещё в середине 1960-х гг. Однако особую актуальность она приобрела только с развитием широкополосного Интернета, сделавшего до-ступным её широкое применение. Национальный институт стандартов и технологий (США) выделяет сегодня следующие особенности облачных вычислений [13, с. 3]:

1. *Самообслуживание по требованию* – пользователь облачных вычислений получает необходимые услуги автоматически в момент обращения без лично контакта с поставщиком услуг.

2. *Широкий доступ* – клиент получает возможность пользоваться облачными вычислениями, используя широчайший спектр периферийного оборудования (ноутбук, мобильный телефон, персональный компьютер и т.д.).

3. *Объединение ресурсов* – облачные вычисления включают в себя одновременно хранение, обработку, память, и полосу пропускания сети, освобождая пользователя от необходимости приобретения, содержания и обновления дорогостоящего оборудования.

4. *Быстрая эластичность* – пользователь облачных вычислений получает мгновенный неограниченный доступ к любым технологическим ресурсам, в любом количестве и в любое удобное для себя время.

5. *Взвешенное обслуживание* – облако автоматически управляет и оптимизирует использование предоставляемых ресурсов, адаптируя их к особенностям запросов пользователей (например, хранение, обработка, пропускная способность, активные учетные записи пользователя и т.д.).

Рыночным субъектам больше не нужно теперь содержать программистов и системных администраторов, создавать сети внутрикорпоративные сети «Extranet», зависеть от дорогостоящего компьютерного оборудования и т.д. Все эти задачи можно решать дистанционно и с минимальными эксплуатационными затратами.

Использование облачных вычислений подразумевает три основные модели обслуживания и четыре модели развёртывания [14]. При этом модели обслуживания различаются по критерию предоставляемых услуг:

1. *SaaS* (*Software as a service*) – предоставление сетевого доступа к облачному программному обеспечению. Все расчёты производятся в облаке, что высвобождает аппаратные ресурсы пользователей, снижая технические требования к их оборудованию. Облако одинаково воспримет запрос с мобильного телефона и персонального компьютера, из web-браузера и через программный интерфейс.

2. *IaaS* (*Infrastructure as a Service*) – предоставление облачной инфраструктуры как услуги. Облако предоставляет пользователю возможности обработки, хранения и передачи информации, в рамках которых он может самостоятельно устанавливать и управлять произвольным программным обеспечением.

3. *PaaS* (*Platform as a Service*) – предоставление облачной платформы для размещения внешних приложений, библиотек и инструментов, поддерживаемых провайдером. Пользователь может управлять также настройкой конфигурации облачной инфраструктуры. При этом потребитель самостоятельно управляет операционными системами, хранением данных и развернутыми запросами.

Модели развёртывания различаются по критериям целевых функций, платности и доступности предоставляемых услуг [13, с. 3]:

1. *Частное облако* (*Private cloud*) – ориентировано на использование одним хозяйствующим субъектом для внутрифирменных целей. Такое облако может принадлежать как самому хозяйствующему субъекту, так и внешнему провайдеру.

2. *Публичное облако* (*Public cloud*) – ориентировано на свободное использование широкой аудиторией пользователей. Такое облако может принадлежать некоммерческим организациям или государственным учреждениям, которые и определяют условия предоставления услуг.

3. *Гибридное облако* (*Hybrid cloud*) – ориентировано на решение технических задач по координации обмена информацией между несколькими облачными инфраструктурами.

4. *Общественное облако* (*Community cloud*) – ориентировано на решение общественно значимых задач. Управлять таким облаком может заинтересованная общественная организация или внешний провайдер.

В результате применения облачных технологий ещё больше снижается трансакционный «входной» барьер для новых участников сетевого рынка. Они освобождаются от затрат на приобретение и эксплуатацию компьютерной техники, решая те же задачи через покупку более дешёвого процессорного времени, дискового пространства и пропускной способности в облачных приложениях. Это намного эффективнее самостоятельного развития «с нуля», так как пользователи получают готовые решения, реализованные с учетом опыта и знаний разработчиков.

Сегодня некоторые авторы предрекают облачным технологиям способность уже в самое ближайшее время «*интенсифицировать динамику текущей длинной волны*» [15, с. 20]. Предпосылки для этого имеются и связаны они с бурным развитием новой формы организации бизнеса – т.н. «*виртуальными организациями*».

Виртуальная организация подразумевает формирование единой управленческо-технологической и информационной среды за счет временного объединения ресурсов различных субъектов [16, с. 177-178]. Такие организации могут не иметь ни юридической оформления, ни постоянного офиса, ни постоянных сотрудников. Они создаются спонтанно под решение какой-либо задачи при наличии платёжеспособного спроса [17]. Первичной средой для создания виртуальных организаций обычно выступают открытые профессиональные сообщества в сети Интернет. В рамках таких сообществ участники размещают информацию о своих профессиональных возможностях, интересах и потребностях, образуют мини-сообщества, рекрутируют сотрудников и объединяются для решения экономических задач.

Появившись в начале 2000-х гг., профессиональные сообщества быстро превратились в глобальных поставщиков трудовых ресурсов для современной сетевой экономики. В качестве примера можно привести французскую социальную сеть «Viadeo», созданную двадцатью энтузиастами в 2004 году. Сегодня «Viadeo Group» владеет 12 глобальными офисами, в которых работает более 400 сотрудников, а пользовательская база сети включает 45 миллионов человек из 226 стран.[****] Аналогичные сети существуют также в России («Профессионалы.ру»), Китае («Tianji»), Индии («ApnaCircle»), Южной Америке («UNYK») и многих других странах.

Использование интернет-рекрутинга в профессиональных сетях коренным образом меняет принципы организации бизнеса. Теперь «*в сетевой экономике основная цель фирмы – не максимизация ценности фирмы, а максимизация ценности фирменной сети*» [9, с. 67]. На первое место выходит скорость реакции исполнителя на сигналы рынка, способность быстро принять и эффективно выполнить полученный заказ. Поэтому в условиях мгновенного общения и передачи большого количества информации, бизнес-процессы уже не ограничиваются формальными рамками экономического влияния отдельных участников рынка и могут вообще не иметь вертикальной структуры. Экономические проекты виртуальных субъектов не привязаны к фирмам-заказчикам, а число участвующих в них сотрудников варьируется в зависимости от стоящих задач.

Виртуальная организация будущего спонтанно создаётся под платёжеспособный спрос и стремительно адаптируется к рыночным изменениям. Уже сегодня наблюдается постепенный переход от экономических понятий «полная занятость» и «постоянная занятость» к новому понятию «проектная занятость». Крайне важен тот факт, что удалённая занятость в сетевой экономике существенно повышает производительность труда и снижает затраты как работодателя, так и сотрудников. Например, в Австралии уже в 2005 году более 29% сотрудников крупных компаний работало вне офиса, тогда как в США этот показатель вообще составил 34% [18, с. 104].

Всё это делает сетевую экономику не только ключевым фактором перехода к новому экономическому циклу, но и первопричиной коренного реформирования институциональ-ной структуры экономики. Не случайно многие авторы определяют сегодня развитие сетевых технологий как «*поворотный момент в истории, не менее значимый, чем промышленная революция*» [19, с. 10].

Налицо парадоксальная ситуация. С одной стороны, происходит очевидная революция в институциональной структуре «новой экономики»: появляются новые формы и методы экономической деятельности. С другой стороны, изменения происходят вне рамок не только традиционной экономической инфраструктуры, но и традиционных социально-экономических институтов.

Мы привыкли, что всякая крупная компания с большими оборотами представляет собой мощную организацию с офисами, сотрудниками и основными фондами. Благодаря облачным технологиям и удалённой занятости в сетевой экономике происходит виртуализация любой экономической деятельности независимо от масштабов, объёмов и географического расположения. Возможно, сегодня влияние облачных технологий пока не очень заметно, но процессы виртуализации бизнеса явно набирают скорость. А с учётом безальтернативности сетевой экономики в качестве основного кандидата на вывод мировой экономики из кризиса актуальность происходящих процессов вообще трудно переоценить.

Дальнейшее развитие сетевой экономики в России и в мире потребует от экономической науки принципиально новых подходов к анализу и интерпретации экономической динамики. На уровне принятия управленческих решений мы неизбежно столкнёмся с необходимостью обобщения накопленного опыта и существенной коррекции всей экономической, образовательной и инфраструктурной политики. Однако другого пути просто не существует. Именно здесь, похоже, скрывается если и не источник роста отечественной экономики предстоящего десятилетия, то, как минимум, источник её устойчивости и конкурентоспособности.

**Примечание:**
\* Дотком – от англ. «dot-com» (рус. «точка-ком») – домен верхнего уровня \*.com, в котором зарегистрирована большая часть сайтов западных компаний.
\*\* Источник: Сайт компании «Никс» (Россия) – http://www.nix.ru/price/index.html.
\*\*\* Закон впервые был сформулирован основателем компании «3Com» Робертом Меткалфом в отношении стандартов локальных вычислительных сетей «Ethernet».
\*\*\*\* Источник: Сайт компании «Viadeo Group». – http://corporate.viadeo.com/en/about-us/the-group.